# Proposed importance of HOCO chemistry: Inefficient formation of $CO_2$ from CO and OH reactions on ice dust

Atsuki Ishibashi[1,2,3], Germán Molpeceres[4], Hiroshi Hidaka[2], Yasuhiro Oba[2], Thanja Lamberts[5,6], Naoki Watanabe[2]

Abstract

With the advent of JWST ice observations, dedicated studies on the formation reactions of detected molecules are becoming increasingly important. One of the most interesting molecules in interstellar ice is $CO_2$. Despite its simplicity, the main formation reaction considered, CO + OH -> $CO_2$ + H through the energetic HOCO* intermediate on ice dust, is subject to uncertainty because it directly competes with the stabilization of HOCO as a final product which is formed through energy dissipation of HOCO* to the water ice. When energy dissipation to the surface is effective during reaction, HOCO can be a dominant product. In this study, we experimentally demonstrate that the major product of the reaction is indeed not $CO_2$, but rather the highly reactive radical HOCO. The HOCO radical can later evolve into $CO_2$ through H-abstraction reactions, but these reactions compete with addition reactions, leading to the formation of carboxylic acids (R-COOH). Our results highlight the importance of HOCO chemistry and encourage further exploration of the chemistry of this radical.

[1] Corresponding author atsukiishibashi@g.ecc.u-tokyo.ac.jp
[2] Institute of Low Temperature Science, Hokkaido University, N19W8, Kita-ku, Sapporo, Hokkaido 060-0819, Japan
[3] Komaba Institute for Science, The University of Tokyo, 4-6-1 Komaba, Meguro, Tokyo 153-8902, Japan
[4] Departamento de Astrofísica Molecular, Instituto de Física Fundamental (IFF-CSIC), Madrid 28006, Spain
[5] Leiden Institute of Chemistry, Gorlaeus Laboratories, Leiden University, P.O. Box 9502, Leiden 2300 RA, The Netherlands
[6] Leiden Observatory, Leiden University, P.O. Box 9513, 2300 RA Leiden, The Netherlands

1. INTRODUCTION

Carbon dioxide ($CO_2$) is one of the most abundant molecules in the ice mantle of dust grains (e.g., Gerakines et al. 1999, Gibb et al. 2000); thus, its formation reactions have been extensively investigated through experiments for a long time. Because the observed solid $CO_2$ abundance cannot be explained by gas-phase synthesis (Hasegawa et al. 1992), chemical formation routes in the solid phase must occur. Since the detection of $CO_2$ production in early works on the UV photolysis of various ice mixtures with CO (e.g., d'Hendecourt et al. 1986, Allamandola et al. 1988, Grim et al. 1989), it has been widely accepted that $CO_2$ can be easily produced by the reaction of CO and OH photofragments from $H_2O$ in ice mixtures. In fact, Watanabe & Kouchi (2002) demonstrated the efficient formation of $CO_2$ from binary CO–$H_2O$ ice. Although an alternative reaction of CO with photoexcited CO*, CO + CO* -> $CO_2$ + O, has been proposed based on the observation of the $CO_2$ product in the UV-photolyzed pure CO solid (Gerakines et al. 1996, Loeffler et al. 2005), the $CO_2$ formation efficiency is negligibly small. The reaction of CO with O atoms has often been examined experimentally (Roser et al. 2001, Madzunkov et al. 2006, Minissale et al. 2013), and in depth computationally (Goumans & Andersson 2010). However, $CO_2$ was detected with very low yield by thermal desorption method which would not be detectable by infrared spectroscopy. Consequently, the following reaction,

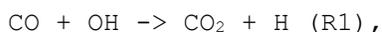
$$CO + OH \rightarrow CO_2 + H \quad (R1),$$

is generally considered the predominant $CO_2$ formation reaction. However, many previous experiments have not discovered information on the thermal reaction of CO + OH on ice surfaces because they observed $CO_2$ products by infrared absorption spectroscopy and/or temperature-programmed desorption methods, which are not surface selective. Moreover, the reactant OH may be energetic upon photodissociation of $H_2O$. Oba et al. (2010a) first observed $CO_2$ formation via the reaction of CO and OH thermalized on an ice surface and detected trace amounts of HOCO on the ice surface using FTIR measurements, clearly indicating the occurrence of

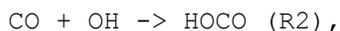
$$CO + OH \rightarrow HOCO \quad (R2),$$

together with reaction (R1). In their experiment, OH was continuously deposited on the ice surface as a gas flowing from a microwave discharge plasma that dissociated $H_2O$ gas. Thus, the gas flow contained other fragments, such as H, O and OH. Under such conditions, HOCO generated via reaction (R2) could further react with these species on the surface, e.g.,

$$\text{HOCO} + \text{H} \rightarrow \text{CO}_2 + \text{H}_2 \text{ and/or HCOOH (R3)},$$
$$\text{HOCO} + \text{OH} \rightarrow \text{CO}_2 + \text{H}_2\text{O and/or H}_2\text{CO}_3 \text{ (R4)}$$

(Yu & Francisco 2008, Yu et al. 2005). The predominance of $CO_2$ products in previous experiments (Watanabe & Kouchi 2002, Oba et al. 2010a, Ioppolo et al. 2011, Yuan et al. 2014, Terwisscha van scheltinga et al. 2022) may have resulted from either energetic OH or an additional reaction. Obviously, to clarify the branching ratio of reactions R1 and R2 on the ice surface relevant to the realistic conditions of dust grains, a new type of experiment is necessary.

The calculated energy diagram for the reaction of CO + OH demonstrates that the formation of $CO_2$ can be achieved through the intermediate HOCO* (e.g., Smith & Zellner 1973, Kudla et al. 1991, Nguyen et al. 2012, Ma et al. 2012):

$$\text{CO} + \text{OH} \rightarrow \text{HOCO}^* \rightarrow \text{CO}_2 + \text{H (R5)}.$$

In the gas phase, the first step of the reaction (R5) has no (or little) activation barrier for the formation of trans-HOCO, while subsequent step for the production of $CO_2$ from trans- and cis-HOCO yields barriers of 8.8 and 2.4 kcal/mol, respectively, with respect to the separated CO and OH (Ma et al. 2012). The gas phase reaction directly produces $CO_2$ once the barriers are overcome either thermally or by quantum tunneling. Recently, the energy diagram for the reaction (R5) on a water cluster was calculated, showing that the barrier heights for the first and second steps are greatly reduced so that it is energetically possible for reaction R5 to complete $CO_2$ formation at 10 K. However, unlike in the gas phase, energy dissipation after the formation of HOCO* on the surface can hinder the overcoming of the barrier of the second step of $CO_2$ formation on the ice surface. Once energy dissipation occurs rapidly, the reaction may end after HOCO formation; i.e., the reaction (R2) can be written as

$$\text{CO} + \text{OH} \rightarrow \text{HOCO}^* \rightarrow \text{HOCO (R2)}.$$

This means that H atom release (R5) and energy dissipation (R2) of HOCO* compete on the surface. And the spontaneous conversion of HOCO stabilized by energy dissipation to $CO_2$ is an endothermic reaction with an energy barrier of ~1 eV (Molpeceres et al. 2023a). In fact, molecular dynamics simulations (Arasa et al. 2013) and coupled quantum chemical and microcanonical models, which parameterized the energy dissipation rate (Molpeceres et al. 2023a), support this scenario rather than $CO_2$ formation. To evaluate the competition between HOCO and $CO_2$ formation, i.e., the role of the ice surface in energy dissipation, an environment needs to be created in which OH is thermalized and reactions of HOCO with the other adsorbates are minimized. However, conducting experiments to identify the reaction products under these conditions is highly challenging with conventional methods, such as the infrared absorption spectroscopy, due to limitations in detection efficiency. The objective of this article is to quantitatively sample the reaction R2, profiting from our newly developed $Cs^+$ ion pickup technique (Ishibashi et al. 2021) for detection of trace species on solid surface.

## 2. EXPERIMENTAL PROCEDURES

The experiments were conducted using our experimental setup for surface reaction experiments, which was equipped with a highly sensitive detection system based on the $Cs^+$ ion pickup method (Ishibashi et al. 2021). An amorphous solid water (ASW) slab (~10 ML) was prepared by background vapor deposition of $H_2O$ onto an Al substrate at 30 K, waiting ~30 minutes for recovery of pressure, followed by the generation of OH after exposing ASW for 1–20 min to UV photons (~$6 \times 10^{12}$ photons $cm^{-2}$ $s^{-1}$) from a conventional deuterium lamp (115–400 nm). Photons from the lamp photodissociate $H_2O$ mainly into H + OH with minor channels, $H_2$ + O and 2H + O (Slanger et al. 1982). However, most of these volatile products on the surface that would be unfavorable for our study, especially the H atom, immediately desorb upon photodissociation at temperatures above approximately 20 K (Al-Halabi & Van Dishoeck 2007, Hama & Watanabe 2013). Therefore, in the present study, the deposition and exposure temperature were set to 30 K. The coverage of OH on the ASW was estimated to be on the order of magnitude of 0.01 or less from the peak signal intensity of the OH radical (See Appendix A). After stopping UV irradiation, the ASW sample was held at 30 K for 30 seconds, followed by cooling to 10 K. Then after a ~20-minute waiting period to neglect the contribution of transient conditions after pre-UV irradiation (e.g., diffusion of H atoms that may remain in very small amounts in ASW and deposition of UV-derived contaminants), CO gas was deposited at a rate of ~$1 \times 10^{11}$ molecules $cm^{-2}$ $s^{-1}$ on the pre-UV-irradiated ASW through a microcapillary plate connected to a gas line different from the $H_2O$ deposition. During CO deposition, the reaction products were monitored. Because the H atoms were negligible, as described previously, reactions resulted from purely CO + OH. Hence, $CO_2$ formation by a subsequent reaction of HOCO with H (reaction R5) and UV (HOCO + hν -> $CO_2$ + H) was neglected in the present experimental system. Considering the low coverage of OH and the amount of CO, gaseous CO was expected to be thermalized to some extent at the surface before colliding with OH. The $Cs^+$ ions were injected at an energy of ~17 eV with a flux of ~1 nA, which enabled us to monitor the reactant and product species (mass X) on the surface with no fragmentation (See Appendix A1 of Ishibashi et al. 2021 in details). The masses of the picked-up species were identified by analysis with a quadrupole mass spectrometer without an ionization cell as a mass number of 133 + X (Kang 2011). With this method, the number of picked-up species was very small, indicating that the chemical composition of the surface did not change; hence, real-time continuous measurements are possible. In fact, Appendix B shows that $Cs^+$ ion bombardment did not affect the reaction system.

## 3. RESULTS AND DISCUSSIONS

### 3.1. Surface reaction products after CO deposition

Fig. 1 shows the pickup spectrum after ~0.1 ML CO deposition in (a) pre-UV-irradiated ASW and (b) blank ASW (i.e., no UV irradiation) cooled to 10 K (Appendix C showed a larger mass region). Peaks that do not change in Figs. 1 (a) and (b) are shown in gray and correspond to $H_2O$ (mass 151 = 133 + 18 m/z) and CO (mass 161 = 133 + 28 m/z). In the following section, only the mass numbers of the adsorbates (unit: u) are shown. The purple peaks represent photoproducts and trace contaminants resulting from pre-UV irradiation (pickup spectrum of pre-UV-irradiated ASW are shown in Fig 5a in Appendix A). The red peaks represent surface reaction products that appear after CO deposition, resulting from the reaction with OH prepared on ASW by pre-UV irradiation. In Fig. 1(a), the pickup signals of masses 44 and 45 u corresponding to $CO_2$ and HOCO, respectively, are clearly detected, while these signals are not detected in Fig. 1(b).

Figure 2 shows the temporal evolution of the pickup intensities for the products during the deposition of CO at 10 K on pre-UV-irradiated ASW. Both mass 45 u signal corresponding to HOCO and mass 44 u signal corresponding to $CO_2$ increase with increasing CO deposition time. These results are in line with the reaction of CO + OH occurring on ASW at 10 K. Note that the Y-axis in Fig. 2 represents the pickup intensities normalized by the $H_2O$ signal intensity in each experiment, in units of "counts" (see Appendix D for further details). Hereafter, all signals used in the analysis were corrected in this manner.

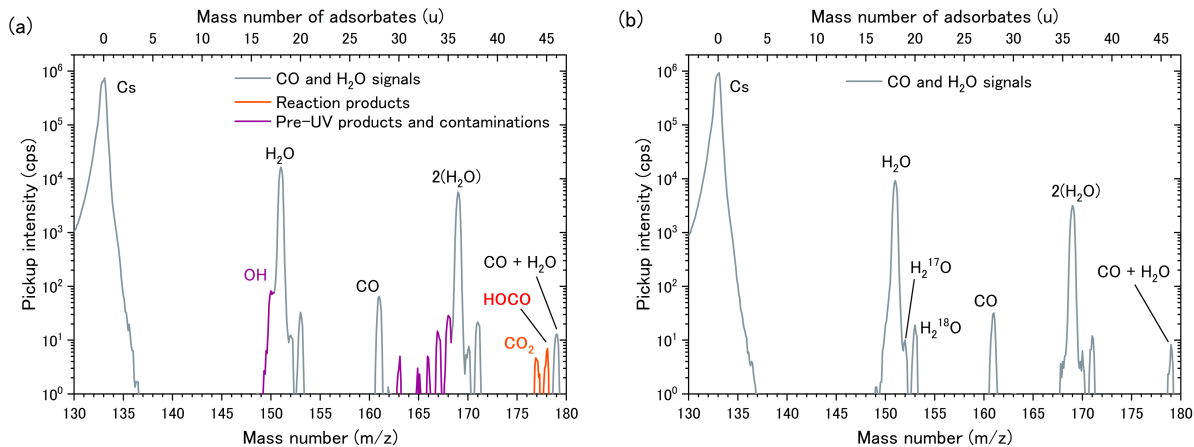

**Figure 1.** Pickup spectra of ∼ 0.1 ML CO deposited on (a) pre-UV irradiated ASW (20 min) and (b) blank ASW at 10 K. Both ASW samples were prepared at 30 K, and the pickup measurements were conducted at 10 K with CO deposition. The unit of "cps (counts per sec)" indicates raw signal values without correction.

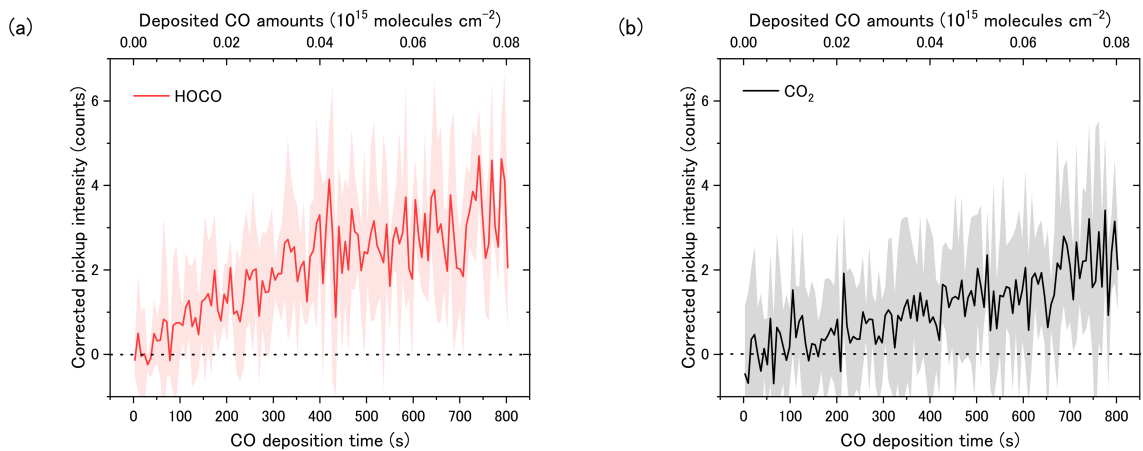

**Figure 2.** Temporal evolution of the signals of surface species obtained at 10 K during CO deposition on the ASW surface bearing OH radicals preproduced via pre-UV irradiation for 20 min. (a) Mass 45 u: HOCO (red). (b) Mass 44 u: $CO_2$ (black). The region after 0 s indicates the duration of CO deposition at a rate of ∼1×10$^{11}$ molecules cm$^{-2}$ s$^{-1}$, as estimated by the increase in the CO infrared band obtained by Fourier transform infrared spectroscopy measurements. The solid lines represent the four different measurements, and the shadows show statistical errors. Vertical dotted lines indicate zero of pickup intensities. The vertical axis is the corrected pickup intensities of products by $H_2O$ signal, in units of "counts". Note that these data are obtained by subtracting the effect of contaminants (see Appendix E).

## 3.2. Estimation of the Branching Ratio

Considering the different pickup efficiencies between HOCO and $CO_2$, the branching ratio of HOCO and $CO_2$ can, in principle, be obtained from the pickup intensity of Fig. 2. However, in the present experimental system, $CO_2$ can be produced not only by reaction R5 but also by the reaction of HOCO with another OH. When $CO_2$ formation from HOCO and OH occurs, its contribution must be removed to determine the branching ratio for the CO + OH reaction alone. To distinguish $CO_2$ formed from each reaction, we consider that the number of OH radicals required to produce $CO_2$ is different in each case, namely, one OH for R5 and two for R2 followed by the reaction of HOCO and OH. The yield of $CO_2$ formed in each reaction should show a different dependence on the initial concentration of OH ($[OH]_0$), and the branching ratio of R2 and R5 is constant for $[OH]_0$. Therefore, we first derive the dependence levels of the reaction yields of HOCO and $CO_2$ on $[OH]_0$ for the reaction system in this experiment. The reaction scheme in the present experiments and the results of the analysis described above are shown in Figure 3. In conclusion, the yield of $CO_2$ formed from CO + OH and HOCO + OH are linear and squared with respect to $[OH]_0$, respectively.

The surface coverage of OH is very low, ≲ 1%; thus, diffusion is necessary for the sequential reaction between HOCO and OH since HOCO is likely not formed on the surface directly next to another OH. These two radicals should not diffuse thermally at 10 K due to their high adsorption energy (Miyazaki et al. 2022, Molpeceres et al. 2023a). Therefore, if a sequential reaction occurs, it should be the reaction between OH and suprathermal HOCO, which can experience transient diffusion with energy dissipation, using the heat of the CO + OH reaction, as follows:

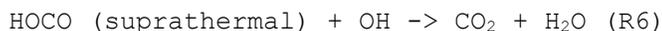

HOCO (suprathermal) + OH -> $CO_2$ + $H_2O$ (R6)

We now derive the rate equations governing our experiment and the mechanisms by which we fit them using our experimental information. For reactions R2 and R5, the deposited CO must be associated with OH on the ASW surface. The following two cases are considered for the association of CO and OH: (i) CO associates with OH during its transient diffusion after landing on the surface before thermalization, and (ii) the thermalized CO and OH adsorbed on ASW associate via their thermal diffusion. On ASW at 10 K, because CO and OH do not experience effective thermal diffusion on a laboratory time scale (< 1 hour) (Kouchi et al. 2021, Miyazaki et al. 2022), most association reactions should occur through process (i). That is, thermalized CO without meeting with OH accumulates on the surface and does not contribute to any reactions. Therefore, under the present deposition conditions, there is no cumulative effect of CO deposited on the rate of reaction with OH. Specifically, the coverage of CO under the present experimental conditions does not affect the reaction, while the flux of CO encountering OH during transient diffusion is essential for determining the rate. Hence, the reaction between CO and OH in the present experimental system can be written using the CO flux, $F_{CO}$ (cm$^{-2}$ s$^{-1}$), and the reaction cross-sections during transient diffusion, $\sigma$ (cm$^2$), similar to a photodissociation process, instead of a second-order kinetic rate constant, k (s$^{-1}$ cm$^2$). A similar argument can be made for the time variation in the concentration of suprathermal HOCO; that is, there is no cumulation effect for the HOCO number density on the surface contributing to the sequential reaction. Therefore, variations in the surface number densities of products X, [X] (cm$^{-2}$), through reactions R2, R5 and R6 in Fig. 3(a) during CO deposition can be described by the following rate equations (a detailed derivation of the equations is shown in Appendix F):

$$\frac{d[\text{HOCO}]_t}{dt} = k_2[\text{CO}_{\text{transient}}]_{\text{const}}[\text{OH}]_t - k_6[\text{HOCO}_{\text{suprathermal}}]_t[\text{OH}]_t$$

$$= \sigma_2 F_{\text{CO}}[\text{OH}]_t - \sigma_6 \sigma_2 F_{\text{CO}}[\text{OH}]_t^2 \quad (1)$$

$$\frac{d[\text{CO}_2]_t}{dt} = k_5[\text{CO}_{\text{transient}}]_{\text{const}}[\text{OH}]_t + \beta_6 k_6[\text{HOCO}_{\text{suprathermal}}]_t[\text{OH}]_t$$

$$= \sigma_5 F_{\text{CO}}[\text{OH}]_t + \beta_6 \sigma_6 \sigma_2 F_{\text{CO}}[\text{OH}]_t^2 \quad (2)$$

where $\beta_6$ corresponds to the branching fraction of CO$_2$ formation in the sequential reaction of suprathermal HOCO + OH, which may lead to products other than CO$_2$, such as H$_2$CO$_3$ (i.e., $0 \leq \beta_6 \leq 1$) (Oba et al. 2010a, b). Specifically, the first term in Equation 1 represents the contribution of HOCO formation by reaction R2, and the second term refers to the consumption by sequential reaction of OH with suprathermal HOCO. The first and second terms in Equation 2 refer to the contributions to CO$_2$ formation from reaction R5 and R6, respectively. Notably, [CO$_{\text{transient}}$]$_{const}$ and $[\text{HOCO}_{\text{suprathermal}}]_t$ are not experimentally measurable due to their short lifetimes, which is very likely between picoseconds and nanoseconds according to the literature on other reactions on water surfaces (Pantaleone et al. 2021, Upadhyay et al. 2022, Molpeceres et al. 2023b, Ferrero et al. 2023).

At the very beginning of CO deposition (t ~ 0 s), Equations 1 and 2 can be regarded as functions of [OH]$_0$. The pickup intensity of surface product X by the Cs$^+$ ions, $I_X$ (t) (counts), can be described as [X] multiplied by its pickup efficiency, $P_X$ (counts cm$^2$); thus, the following equations can be written as follows:

$$\frac{d}{dt}I_{HOCO\,(t\sim 0)} = \sigma_2 F_{CO}\frac{P_{HOCO}}{P_{OH}}I_{OH\,(t=0)} - \sigma_6\sigma_2 F_{CO}\frac{P_{HOCO}}{P_{OH}^2}I_{OH\,(t=0)}^2$$

$$= A \times I_{OH\,(t=0)} - C \times I_{OH\,(t=0)}^2 \quad (3)$$

$$\frac{d}{dt}I_{CO2\,(t\sim 0)} = \sigma_5 F_{CO}\frac{P_{CO2}}{P_{OH}}I_{OH\,(t=0)} + \beta_6\sigma_6\sigma_2 F_{CO}\frac{P_{CO2}}{P_{OH}^2}I_{OH\,(t=0)}^2$$

$$= B \times I_{OH\,(t=0)} + D \times I_{OH\,(t=0)}^2 \quad (4)$$

In equation (3) and (4), $\frac{d}{dt}I_{X\,(t\sim 0)}$ refers to the slope of the time variation for $I_X$ (t) at approximately t = 0, which is obtained from the linear increasing region of $I_X$ (t) in Fig. 2 (t = 0–600 s). Hence, the CO$_2$ production reactions can be distinguished by analyzing $\frac{d}{dt}I_{X\,(t\sim 0)}$ versus the initial pickup intensity of OH before CO deposition ($I_{OH\,(t=0)}$), which corresponds to [OH]$_0$.

Figure 3 shows the dependence levels of (b) $\frac{d}{dt}I_{HOCO\,(t=0-600)}$ and (c) $\frac{d}{dt}I_{CO2\,(t=0-600)}$ on $I_{OH\,(t=0)}$ from each experiment where different [OH]$_0$ samples are prepared by controlling the pre-UV irradiation time (see Fig 5b in Appendix A for pre-UV time dependence of $I_{OH\,(t=0)}$). From the fitting of Figs. 3(b) and (c) using Equations 3 and 4, respectively, we obtain the following values: A = (7.81 ± 0.27) × 10$^{-5}$ s$^{-1}$, B = (3.27 ± 1.80) × 10$^{-6}$ s$^{-1}$, C = (2.60 ± 0.26) × 10$^{-7}$ s$^{-1}$ counts$^{-1}$, and D = (1.49 ± 0.18) × 10$^{-7}$ s$^{-1}$ counts$^{-1}$. From Equations (3) and (4), the linear term of the fitting parameters, A and B, yields the branching ratio of CO + OH, $\frac{\sigma_2}{\sigma_5} = \frac{A}{B} \div P_{HOCO/CO2}$. Where $P_{HOCO/CO2}$ is the relative pickup efficiency between HOCO and CO$_2$. Then, $P_{HOCO/CO2}$ can be expressed as $\frac{C}{D} \times \beta_6$ from the fitting parameters, C and D, referred to the second terms in Equations 3 and 4. Therefore, if the branching value $\beta_6$ for suprathermal HOCO + OH is known, $P_{HOCO/CO2}$ and $\frac{\sigma_2}{\sigma_5}$ can be accurately obtained. However, it is difficult to estimate $\beta_6$ from this experiment because the pickup signal of H$_2$CO$_3$ (mass 62 u) overlaps with multiple pickup signals of CO$_2$ and H$_2$O (mass 44 + 18 u). Assuming $\beta_6$ = 1.0 (i.e., suprathermal HOCO + OH only leads to CO$_2$ formation), $P_{HOCO/CO2}$ has an upper limit of 1.7 ± 0.3. Hence, $\frac{\sigma_2}{\sigma_5}$ is obtained as a lower limit of 13.6 ± 7.9. The large error of $\frac{\sigma_2}{\sigma_5}$ originates from errors of A and B values derived from the fitting uncertainties. Relatively small value of $\sigma_5$ leads to the large error of $\frac{\sigma_2}{\sigma_5}$. The HOCO formation ratio is at least ~93% (85-96%). In reality, $\beta_6$ would be less than 1.0, so its formation ratio would be much larger. In other words, this finding suggests that the contribution of $\sigma_5$ (i.e., direct formation of $CO_2$ from CO + OH on ASW) is very minor.[7]

(a)

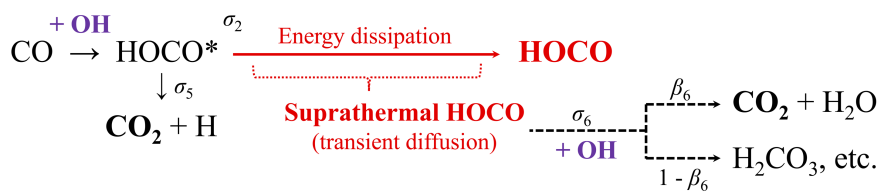

(b)

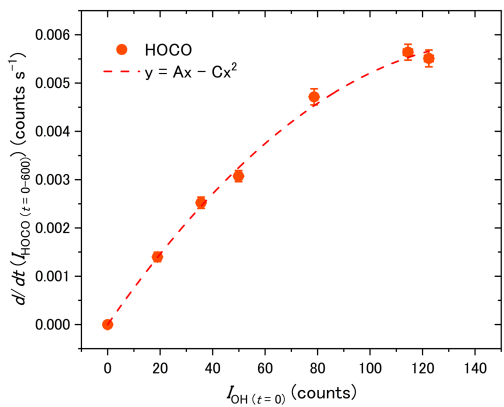

(c)

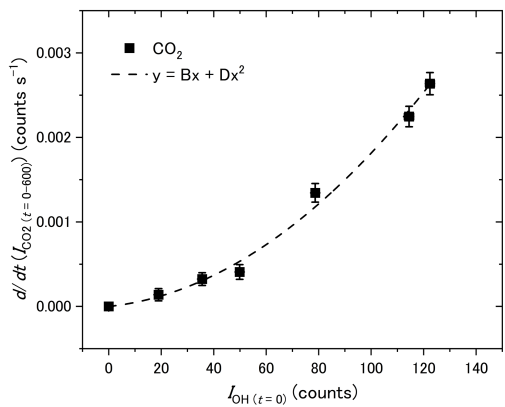

---

[7] We also investigated CO deposition experiments at 40 K, although the thermal diffusion of reactants and products makes detailed analysis difficult. At 40 K, a ratio of pickup intensities for HOCO and $CO_2$ shows roughly the same as 10 K, indicating that HOCO may still be in rich when the contribution of sequential reactions is similar (see Figure 7).

**Figure 3.** (a) Reaction scheme that occurs in the present experiment. (b, c) [OH]$_0$ dependence of the formation slopes for the reaction products (b) HOCO and (c) CO$_2$ at t ~ 0 s, corresponding to the linear growth region (t = 0–600 s), during CO deposition. Each plot and error bar were obtained from the results of four experiments. The pre-UV irradiation times were 1, 2, 3, 5, 10, and 20 min. (b) and (c) were fitted by y = Ax – Cx$^2$ and y = Bx + Dx$^2$, respectively. Fitting results; A = $\sigma_2 F_{CO} \frac{P_{HOCO}}{P_{OH}}$ = (7.81 ± 0.27) × 10$^{-5}$, B = $\sigma_5 F_{CO} \frac{P_{CO2}}{P_{OH}}$ = (3.27 ± 1.80) × 10$^{-6}$, C = $\sigma_6 \sigma_2 F_{CO} \frac{P_{HOCO}}{P_{OH}^2}$ = (2.60 ± 0.26) × 10$^{-7}$, and D = $\beta_6 \sigma_6 \sigma_2 F_{CO} \frac{P_{CO2}}{P_{OH}^2}$ = (1.49 ± 0.18) × 10$^{-7}$. The units of A and B are s$^{-1}$ and of C and D are s$^{-1}$ counts$^{-1}$.

## 4. Astrophysical Implications

Using the novel, in situ, nondestructive Cs$^+$ ion pickup technique, we constrain the fraction of CO$_2$ coming from reaction R5 to be more than one order of magnitude lower than the amount of HOCO formed by reaction R2. This result is the main finding of our work that aligns with recent theoretical predictions (Arasa et al. 2013, Molpeceres et al. 2023a). Here, we argue that a new evaluation of interstellar CO$_2$ formation pathways, as implemented in astrochemical models, is needed.

The formation of CO$_2$ has been extensively studied from an energetic standpoint. Initially, the reaction between CO and O has been considered, with reported activation energies of either ~300 K (Roser et al. 2001) or higher, between ~500-800 K (Minissale et al. 2013). These activation energies could make the CO + O reaction an important route to CO$_2$ compared to the title reaction. However, theoretical studies predict much higher activation energies (Talbi et al. 2006, Goumans et al. 2008), highlighting the need for further studies to elucidate the significance of the CO + O reaction.

Focusing on the route studied in this work, the traditional pathway for the title reaction, CO + OH -> CO$_2$ + H, has been reported in astrochemical literature to have very low activation energies, in the range of 80-500 K (Noble et al. 2011, Oba et al. 2010a). Recently, Molpeceres et al. 2023a showed that these values align with the formation of HOCO rather than of CO$_2$, and a simple correction by substituting the products of the CO + OH reaction is sufficient to update the models. The reaction HOCO -> CO$_2$ + H has a very high barrier of >15000 K, making it unlikely to be overcome thermally or even with external energy input Molpeceres at al. 2023a, finishing the reaction at HOCO. The authors of Molpeceres at al. 2023a concluded that the formation of $CO_2$ observed in previous experimental literature must be explained by additional surface reactions in the experimental setup. This explanation is consistent with those experiments and aligns the theoretical results of Molpeceres at al. 2023a with the results reported here.

The main differences between our experiments and the conditions found in the cold ISM are (I) the surface coverage of OH radicals, (II) the accretion rate of CO molecules, and (III) the thermalization scenario for CO molecules landing on the surface. In reality, the surface coverage of OH radicals on ISM grains is much lower, especially in molecular clouds with high extinction ($A_v$), which suppresses the main channel for producing suprathermal OH (translationally excited) by the photodissociation of $H_2O$ on ice. In addition, the accretion rate of CO molecules under ISM conditions is lower, and accreted CO molecules on ice rarely reach the reaction partner; therefore, they thermalize on the surface before the reaction, i.e., they do not participate in transient diffusion. A subsequent reaction step requires thermal CO diffusion to meet an OH partner, a process that competes with CO hydrogenation (Watanabe & Kouchi 2002b). The third alternative involves the formation of nonthermal OH radicals, i.e., O + H -> OH, which may still react with CO, a process that is possible in molecular clouds. When the information revealed by our experiments is translated into ISM conditions, the competition between reaction R5 (direct $CO_2$ formation) and a two-step mechanism such as reactions R2 + R3 clearly favors the latter.
We recommend future modeling endeavors in light of JWST ice observations to incorporate (at least) R2 and R3 instead of R1, as given in Fig 4. Moreover, we also suggest to include nonthermal mechanisms exemplified by reaction R6 on top of the thermal mechanisms. The finding that HOCO is the main product of the reaction of CO and OH indicates the need for an accurate determination of the reaction channels and relative branching ratios for general follow-up reactions with HOCO:

$$HOCO + R \rightarrow RH + CO_2$$
$$HOCO + R \rightarrow R\text{-}COOH,$$

These reactions include the formation of carboxylic acids (e.g., formic acid and carbonic acid through radical addition to HOCO) and hydrogen abstraction to form $CO_2$, placing HOCO at the forefront of the field of interstellar ice chemistry. We note that H-abstraction reactions must compete with radical addition reactions, and the formation of acids from the abovementioned radical couplings, such as acetic acid ($CH_3COOH$), carbamic acid ($NH_2COOH$), or the recently detected carbonic acid ($H_2CO_3$) (Sanz-novo et al. 2023), must also be considered.

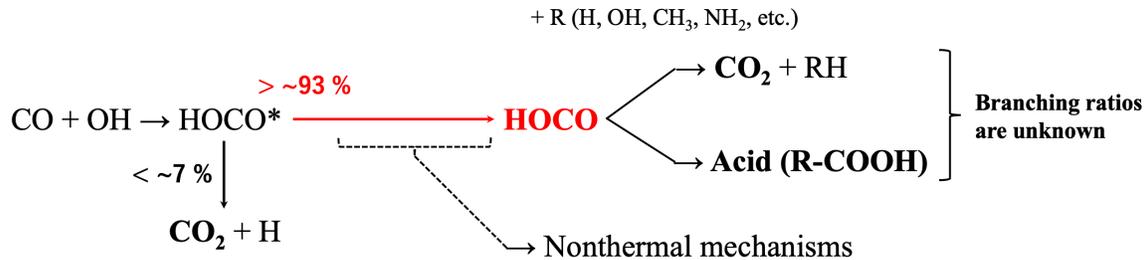

**Figure 4.** Chemical reactions on the actual dust surfaces to be included in future modeling.

In light of the information on the thermal diffusion of radicals on interstellar surfaces at low temperatures (~10-20 K), R can be considered to be hydrogen, oxygen, or nitrogen atoms that can diffuse on the surface (Hama et al. 2012, Ásgeirsson et al. 2017, Minissale et al. 2016, Zaverkin et al. 2022). In addition to atoms, some polyatomic radicals may diffuse on ASW in that temperature window, such as $CH_2$ or $CH_3$, but here, we focus on species with an already determined diffusion energy. Between 20 and 30 K, other radicals can react with HOCO, for example, the C atom (Tsuge et al. 2023). Above 30 K, the OH radical (and possibly other similar radicals, such as $NH_2$, HS, HCO, and $CH_3O$) can undergo thermal diffusion (Miyazaki et al. 2022, Tsuge & Watanabe 2021), contributing to the formation yields of $CO_2$. In other words, H atom abstraction reactions by these radicals from HOCO promote $CO_2$ formation, but addition reactions inhibit $CO_2$ formation. The study of these reactions will be the subject of a subsequent study.

Finally, it is important to note that the discussion in this paragraph assumes the thermal diffusion of radicals, but in reality, nonthermal mechanisms (Shingledecker et al. 2018, Jin & Garrod 2020, Garrod et al. 2022) may occur simultaneously, lifting the temperature constraints discussed above for at least a fraction of the chemical routes under discussion (i.e., suprathermal HOCO + OH). A similar phenomenon has been reported in our previous work in which $CH_3O$ radicals formed from $CH_3OH+OH$ diffuse at 10 K by nonthermal mechanisms (Ishibashi et al. 2024). This phenomenon needs further investigation because it is important for the formation of complex organic molecules in cold environments.

## 5. Acknowledgments

This work was supported by JSPS Grant-in-Aid for Scientific Research JP22H00159 and JP24K17107. The authors thank the staff of Technical Division at Institute of Low Temperature Science for making various experimental devices.

## Appendix

### A. Pre-UV-Irradiated Amorphous Solid Water Samples before CO Deposition

Figure 5 (a) shows the pickup spectrum of pre-UV-irradiated ASW (20 min). Pre-UV irradiation was performed at 30 K and measured at 10 K. Main product was OH radical at mass 17, and other minor photoproducts $HO_2$ (mass 33) and $H_2O_2$ (mass 34) were detected. As the signal intensity of OH is ~1 % of that of $H_2O$ in the condition of 20 min pre-UV-irradiated ASW, the surface coverage of OH is considered to be around 0.01 ML, assuming that their pickup efficiencies are the same (Ishibashi et al. 2024). Tiny peaks appearing at masses 30, 44 and 45 were contaminants due to UV irradiation. A signal of mass 32 was a photoproduct $O_2$ and/or the contaminants. These masses 44 and 45 contributions are subtracted from the $CO_2$ and HOCO formed in the surface reactions. Figure 5 (b) indicates the pre-UV irradiation time dependence of pickup intensity of OH ($I_{OH\,(t=0)}$), which are corrected by $H_2O$ intensity by a method shown in Appendix C. For Pre-UV irradiation times shorter than 20 min, the concentration of OH is even smaller than ~0.01 ML. These values of $I_{OH\,(t=0)}$ are used for horizontal axis of Fig 3 (b) and (c).

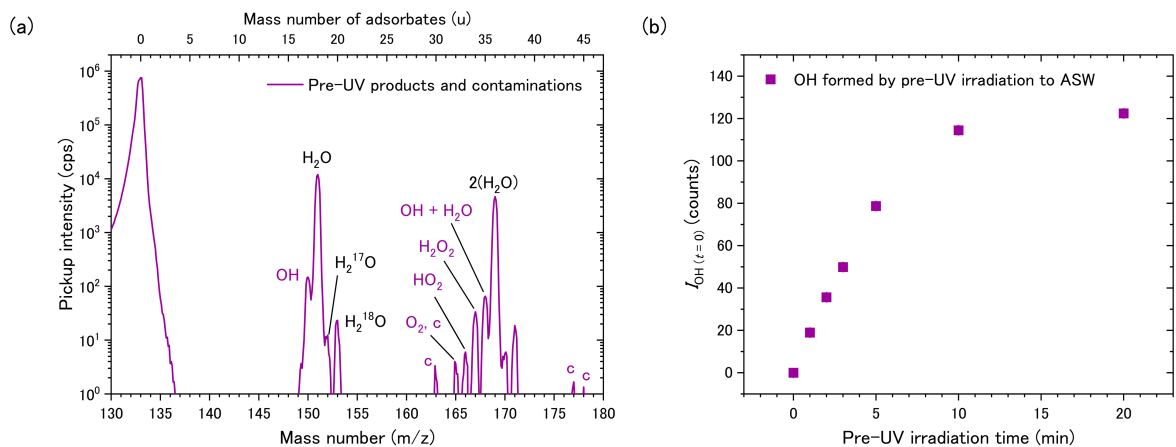

**Figure 5.** (a) Pickup spectrum of pre-UV-irradiated ASW (20 min). "c" refers to trace amounts of contaminants produced by pre-UV irradiation. (b) Pre-UV irradiation time dependence of pickup intensity for OH radical of photoproducts. Pre-UV irradiated time are 1, 2, 3, 5, 10, and 20 min.

B. Absence of an Effect of $Cs^+$ Ion Irradiation on Surface Reaction in the Present Experimental System

First of all, it should be noted that our monitoring reaction leading to HOCO production is barrier-less radical reaction as described in the text. Basically, reaction should proceed without the assistance of $Cs^+$ ion collision. Nevertheless, we here evaluate the possible effects of $Cs^+$ ion on the reaction which we monitored. There are two possible scenarios for the enhancement of reaction by $Cs^+$ ion collision. In the first case, the products formed by assistance of the $Cs^+$ ion collision and remaining on ASW are detected by the subsequent $Cs^+$ ion pickup. In this scenario, the amounts of detected products should increase with the fluence of $Cs^+$ ions. Fig. 6 shows the pickup intensities of the products obtained in the CO deposition experiments. The solid lines represent temporal variation for intensities of the products obtained by continuous $Cs^+$ ion irradiation, while single data point in each figure represents the pickup intensity obtained by instantaneous $Cs^+$ irradiation conducted only at around 600 s of CO deposition time in a different experiment from the continuous monitoring. The pickup signal intensities at 600 s for the different kinds of measurements agreed regardless of the total $Cs^+$ ion irradiation time. This indicates that the enhancement of the reaction by $Cs^+$ ion process does not exist. In the second case, a single $Cs^+$ ion assists reaction and picks-up its product simultaneously. That is, the single $Cs^+$ ion hits CO or OH at the site where CO and OH adsorb nearby to produce HOCO and pickup it simultaneously. Obviously, this second case is more unlikely than the first scenario because this process consists of two elementary phenomena, collisional excitation of reactant and picking up the product. Nevertheless, to exclude the second scenario experimentally, we monitored the HOCO intensities during CO exposure of sample at 40 K where the residence time of CO on ASW should be very short. Figure 7 shows the intensities of parent CO and products of HOCO and CO2 during CO deposition. The HOCO and $CO_2$ were detected with the comparable amounts to those at 10 K despite the parent CO intensities were significant smaller. This is the direct evidence that HOCO were produced promptly at collision between CO and OH within the short residence time of CO in which $Cs^+$ ions have almost no chance to interact with CO and thus the detected HOCO intensities result from the HOCO products on the surface without assistance of $Cs^+$ ion for CO + OH. That is, the second scenario should be negligible.

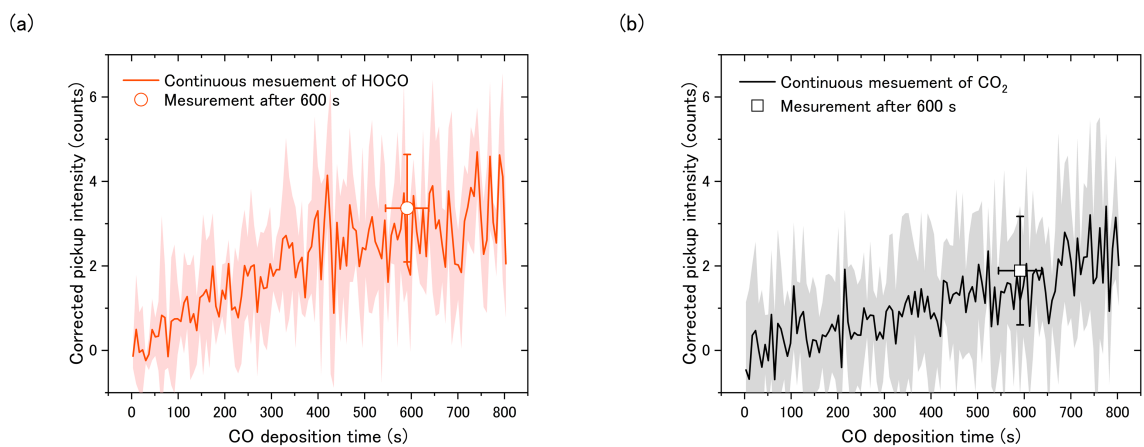

**Figure 6.** Variations in the temporal evolution of HOCO (red) and $CO_2$ (black) signals during CO deposition using different $Cs^+$ ion irradiation procedures: (solid lines) continuous $Cs^+$ irradiation and (symbols) experiments with $Cs^+$ irradiation at ~600 s. Since these values agree, the contribution of $Cs^+$ ion irradiation (0-600s) to the chemical reactions in this experimental system is negligible.

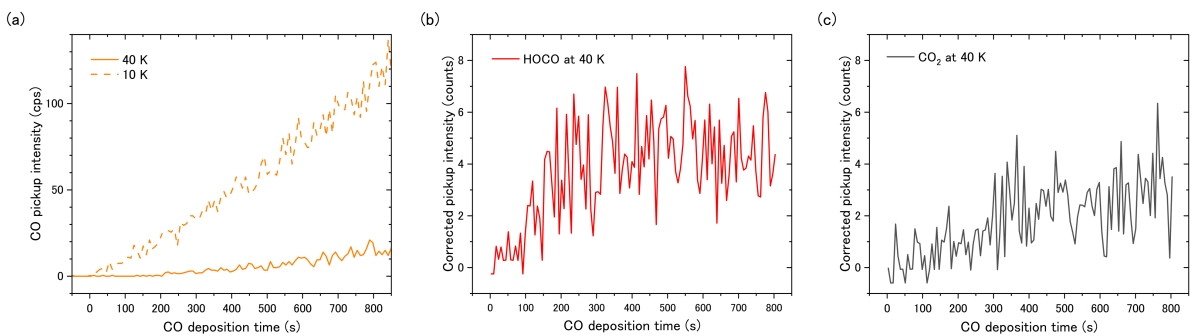

**Figure 7.** Temporal variation of pickup intensities for (a) CO, (b) HOCO, and (c) $CO_2$ during exposure of pre-UV 20 min irradiated ASW to CO gas at 40 K. For (a) CO, results at 10 K are also plotted in the same scale of Y axis. The CO gas flow rate was the same as that for the experiment at 10 K described in the text. Y-axes of (b) and (c) are the same scale as those for Fig. 2 and thus intensities of HOCO and $CO_2$ at 40 K can be directly compared with those at 10 K in Fig. 2. At the beginning of CO deposition, the increase rate of HOCO and CO2 at 40 K is larger than that at 10 K, and signal growth appears to have stopped at ~300 s. This may be due to active thermal diffusion of CO at 40 K regardless of its short residence time, which increases the rate of association with OH radicals which hardly diffuse thermally at 40 K in the laboratory time scale [Miyazaki et al. 2022].

C. Pickup spectrum of CO deposition to pre-UV 20 min ASW surface in mass range of 180-230 m/z

Figure 8 shows the pickup spectrum of pre-UV irradiated (20 min) ASW after CO deposition in mass range of 180-230, which is same condition to Fig. 1a. No products other than those discussed in the main text were detected.

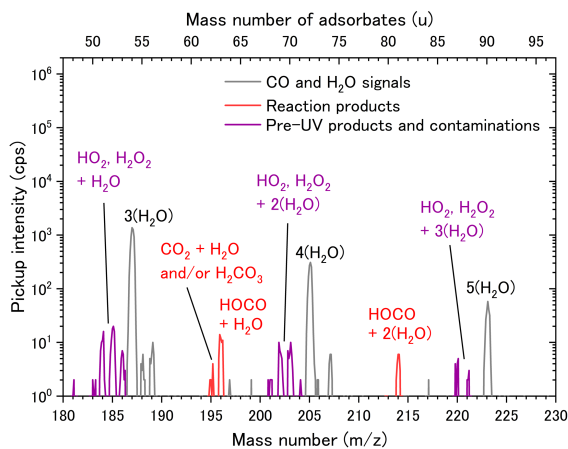

Figure 8. Pickup spectrum of ~ 0.1 ML CO deposited on pre-UV irradiated ASW (20 min) at 10 K in mass range of 180-230 m/z, which is same condition to Fig. 1a.

D. Correction of Pickup Signals for Quantitative Analysis

In the $Cs^+$ ion pickup method, the absolute intensities of the pick-up ions can change in each experiment even under the same experimental conditions. Hence, for quantitative evaluation, the pickup signals of all species in each experiment need to be corrected as a relative intensity to the $H_2O$ pickup signal intensity. As shown in Equation A1, the corrected pickup intensity of adsorbate X, $I_X(t)$, is obtained by dividing the raw pickup intensity of adsorbate X, $I_{rawX}(t)$, by the raw pickup intensity of $H_2O$, $I_{rawH2O}(t)$, obtained in the same experiment and multiplying by $10^4$, which is the standard pickup intensity of $H_2O$. Specifically, the pickup intensities of the adsorbates are corrected by assuming that the $H_2O$ signal is obtained at a pickup intensity of $10^4$ in all experiments.

$$I_X(t) \text{ (counts)} = I_{rawX}(t) \text{ (cps)} \div I_{rawH2O}(t) \text{ (cps)} \times 10^4 \text{ (counts)} \quad (C1)$$

E. Subtracting Contamination Effects

Contamination such as $CO_2$ may supply due to the accumulation of residual gases over time and due to UV irradiation inside the chamber. Therefore, their contribution must be estimated and subtracted in this experiment. The data shown in Fig. 2 are obtained by subtracting these two contributions. Fig. 9 a and b showed the time evolution of mass 45 and 44 signals for CO deposition to pre-UV 20 min ASW before subtracting contamination effects, respectively, which are corresponding to Fig. 1a. The non-zero signal intensity at t=0 indicates the contamination caused by the interaction of UV and chamber wall. Fig. 9 c shows the dependence of the signal from contamination in Mass 44 and 45 on pre-UV irradiation time which were measured 20 minutes after pre-UV was stopped (just before the start of CO deposition), which are corresponding to Fig. 5a. These contributions increase with pre-UV irradiation time. These values were subtracted from Fig. 9 a and b, respectively, as a baseline with no time variation. Additionally, the effect of residual gas during CO deposition was estimated using the results of the blank (no pre-UV) experiments (Fig. d and e), which are corresponding to Fig. 1b. These were subtracted from Fig. 9 a and b, respectively, as time-dependent values.

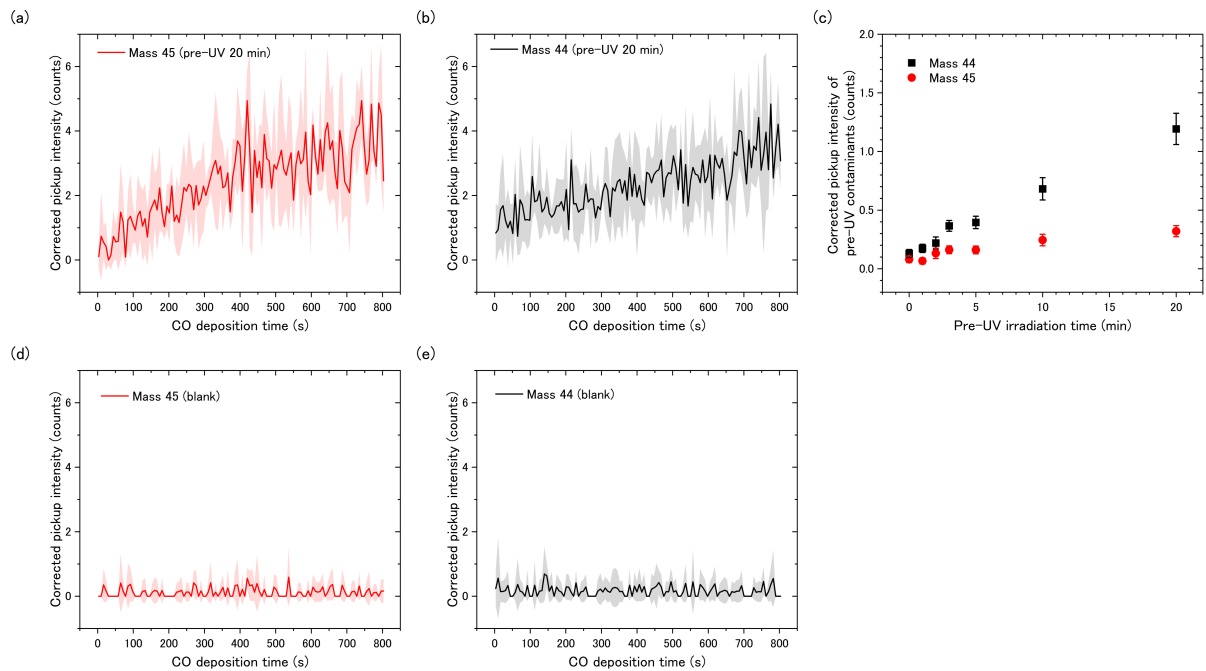

Figure 9. (a) Mass 45 and (b) mass 44 signals of pre-UV 20 min sample during CO deposition before subtracting contamination. (c) pre-UV irradiation time dependences of contaminants produced by UV. (d) Mass 45 and (e) mass 44 signals of blank (no pre-UV) experiments during CO deposition.

F. Derivation of Rate Equations for Reactions Occurring in the Present Experiment

The reactions for which we derive the rate equations in the main text are as follows:

$$CO_T + OH \rightarrow HOCO^* \rightarrow HOCO_T \rightarrow HOCO \quad (F2)$$
$$CO_T + OH \rightarrow HOCO^* \rightarrow CO_2 + H \quad (F3)$$
$$HOCO_T + OH \rightarrow CO_2 + H_2O \quad (F4)$$
$$HOCO_T + OH \rightarrow H_2CO_3 \quad (F5)$$

The subscript "T" indicates a molecule undergoing transient diffusion using adsorption or reaction energy. As mentioned in the main text, under the conditions of present experiments, fully thermalized CO and HOCO would not participate in the reaction because they cannot diffuse well at 10 K.

Variations in the surface number density of products through reactions F2–F5 can be described using the second-order kinetics rate constants, k, by the following rate equations, as mentioned in the main text:

$$\frac{d[HOCO]_t}{dt} = k_2[CO_T]_{const}[OH]_t - k_6[HOCO_T]_t[OH]_t \quad (F6)$$

$$\frac{d[CO_2]_t}{dt} = k_5[CO_T]_{const}[OH]_t + \beta_6 k_6[HOCO_T]_t[OH]_t \quad (F7)$$

$[CO_T]$ is expressed as the flux of CO deposition, $F_{CO}$, (cm$^{-2}$ s$^{-1}$) × lifetime of $CO_T$ (s); therefore, even if the lifetime is 1 ms (generally n-ps, as mentioned in the main text), the surface concentration is $10^8$ cm$^{-2}$, which is below the detection limit in the present experiment. Hence, the detected CO and HOCO pickup signals only reflect information on the thermalized species.

As mentioned in the main text, the reaction terms for $CO_T$ and OH in Equations F6 and F7 can be rewritten as equations for first-order reactions such as photodissociation reactions in UV irradiation experiments, as shown in the following equation:

$$k_{2 \text{ or } 5} \text{ (cm}^2 \text{ s}^{-1}) \times [CO_T]_{const} \text{ (cm}^{-2}) = \sigma_{2 \text{ or } 5} \text{ (cm}^2) \times F_{CO} \text{ (cm}^{-2} \text{ s}^{-1}) \quad (F8).$$

Even in the case of sequential reactions with transient diffusion $HOCO_T$ (i.e., $HOCO_T$ + OH), thermalized HOCO does not contribute to the reaction; thus, it should be approximately expressed as follows in the same manner as Equation F8:

$$k_6 \times [HOCO_T]_t = \sigma_6 \times F_{HOCO} \text{ (t)} \quad (F9)$$

Note, however, that the flux of HOCO, $F_{HOCO}$, is not constant with time since $[HOCO_T]_t$ is not constant. Additionally, $HOCO_T$ is not detectable for the same reasons as $CO_T$. $F_{HOCO}(t)$ is equal to the time derivative of HOCO in the absence of consumption by the sequential reaction, suggesting that $F_{HOCO}(t)$ is equal to the formation term (first term) in Equation F6. Thus, HOCO can be written as follows:

$$F_{HOCO} \text{ (t)} = \sigma_2 \times F_{CO} \times [OH]_t. \quad (D10)$$

Equations 1 and 2 in the main text can be obtained by substituting Equations F9 and F10 into Equations F6 and F7.